\documentclass[slac_one]{revtex4}
\usepackage{graphicx}
\usepackage{fancyhdr}
\begin{document}

\title{Inclusive W and Z production with CMS at LHC startup} 

%

\author{R. Paramatti\footnote{on behalf of the CMS Collaboration}}
\affiliation{INFN Rome, P.le Aldo Moro 2, 00185 Rome, Italy}

\begin{abstract}
We report on potential for measurement of inclusive W and Z boson
 production cross section using initial data from the LHC. We have
 designed W and Z triggers, selection algorithms, and event reconstruction
 techniques for both muon and electron decay modes, for low luminosity
 operation of the LHC integrating up to about 10 pb$^{-1}$. Initial
 calibrations and alignment accuracies are assumed. While the accuracy
 of the cross section extracted will be dominated by the integral luminosity
 measurement, ratios of W and Z production, and asymmetry distributions will
 be important early measurements from LHC.
\end{abstract}

\maketitle

\thispagestyle{fancy}


\section{INTRODUCTION} 
An early measurement of the inclusive W and Z production cross section in leptonic decay
 channels is presented, assuming 10 pb$^{-1}$ data at LHC.
Signature of high transverse momentum leptons from W and Z decays is very distinctive in
the environment of hadron collisions. As such, the decays of W and Z bosons into leptons provide
a clean experimental measurement of their production rate.
The $W \rightarrow l\nu$ cross section can be calculated using the following formula (a similar formula
can be used for the Z cross section):
\begin{equation}
\sigma_W \cdot BR(W \rightarrow l\nu) = \frac{N_W^{pass} - N_W^{bkgd}}{A_W \cdot \epsilon_W \cdot \int L dt}
\end{equation}
where $N_W^{pass}$ is the number of candidates selected from the data, $N_W^{bkgd}$ represents the expected
background events and $A_W$ is the acceptance defined as the fraction of these decays satisfying
the geometric constraints of the detector and the kinematic constraints of the imposed selection
criteria. The $\epsilon_W$ is the selection efficiency for W decays falling within the acceptance and $\int L dt$
is the integrated luminosity.

\section{EVENT SELECTION}

\begin{figure*}[t]
\centering
\includegraphics[width=80mm]{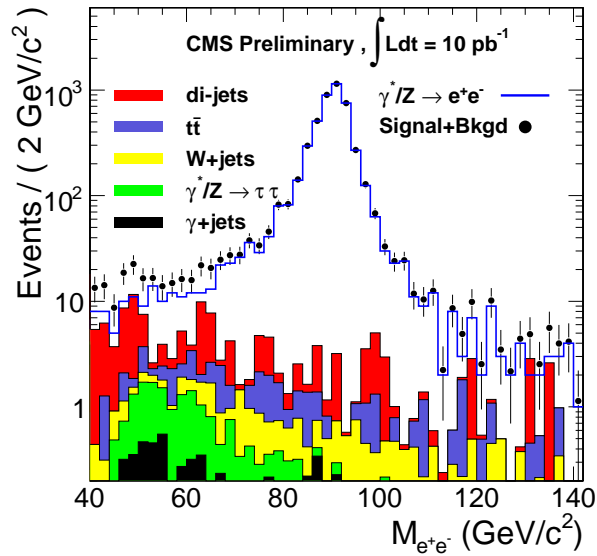}
\caption{The $M_{ee}$ for the $Z \rightarrow ee$ signal together with the considered backgrounds after
 all selection cuts but the invariant mass one.} \label{ZeeInvMass}
\end{figure*}

The $W \rightarrow e\nu$ and $\gamma^*/Z \rightarrow e^+e^-$ ($Z\rightarrow ee$ in the following) samples are selected 
from events that pass the single isolated electron High Level Trigger~\cite{HLT}. 
We require one (for $W \rightarrow e\nu$) or two (for $Z \rightarrow ee$)
high-$p_T$ electrons formed from the association of a high $E_T$ supercluster in the $PbWO_4$ crystal
electromagnetic calorimeter (ECAL) and a high-$p_T$ GSF track in the Tracker. An ECAL
supercluster gathers the energy deposited in a region around the main energy cluster in an
attempt to recover most of the energy of Bremsstrahlung photons emitted along the electron
trajectory. The momentum of a GSF track is fitted along its trajectory using a Gaussian-Sum
Filter algorithm (GSF) dealing with the possible emission of hard Bremsstrahlung photons in
the scattering layers of the Tracker.
The electron(s) should fall within the ECAL fiducial region ($|\eta| < $ 2.5, excluding the Barrel-Endcap 
transition region). The ECAL supercluster(s) should have a transverse energy $E_T > $ 20.0 (30.0) GeV for 
$Z \rightarrow ee$ ($W \rightarrow e\nu$).
Since the electrons from the Z and W decays are isolated, we demand low charged particle activity in a
cone around each electron candidates.
The reconstructed $M_{e,e}$ distribution for the signal and the various backgrounds for events passing
the $Z \rightarrow ee$ selection is shown in Fig.~\ref{ZeeInvMass}. An invariant mass cut of
70-110 GeV is applied. The background after selection is negligible.

\begin{figure*}[h]
\centering
\includegraphics[width=80mm]{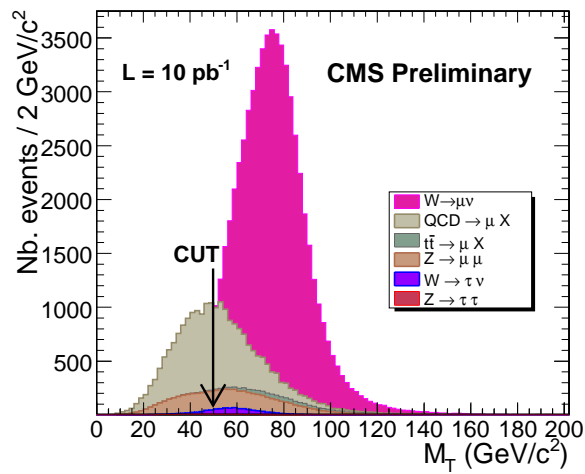}
\caption{Reconstructed transverse mass $M_T$ of W candidates. All $W \rightarrow \mu\nu$ selection cuts but the one shown 
in the plot have been applied.} \label{WmunuMet}
\end{figure*}

In order to select the $Z \rightarrow \mu^+\mu^-$ sample, we require two
isolated muons with tracks reconstructed from hits in both the tracker system and the muon
chambers. The muons must satisfy a cut on the transverse momentum: $p_T >$~20~GeV. 
The isolation criteria requires the $p_T$ sum of all tracks in a cone around
the muon direction to be less than 3 GeV. The invariant mass of the $ \mu^+\mu^-$ pair 
must be greater than 40 GeV.\\
Fig.~\ref{WmunuMet} shows the reconstructed transverse mass, $M_T$, of the W system after $W \rightarrow \mu \nu$ selection
cuts. The W system is built in the plane transverse to the beam by combining the measured
muon and the missing transverse energy in the event. The latter is interpreted as a measurement
of the transverse momentum of the undetected neutrino. Muons must satisfy the cuts:
$p_T >$ 25 GeV and $|\eta| <$ 2 and be isolated. The isolation criteria for $W \rightarrow \mu \nu$ requires the $p_T$ sum
of all tracks cone around the muon direction, normalized to the muon $p_T$, to be
less than 0.09. The figure shows that the QCD background is largely suppressed with the cut $M_T >$ 50 GeV.


A full description of the selection is described in \cite{PASele} for electron channels and in \cite{PASmuon} for muon channels.


\section{TAG AND PROBE METHOD}
\begin{figure*}[h]
\centering
\includegraphics[width=75mm]{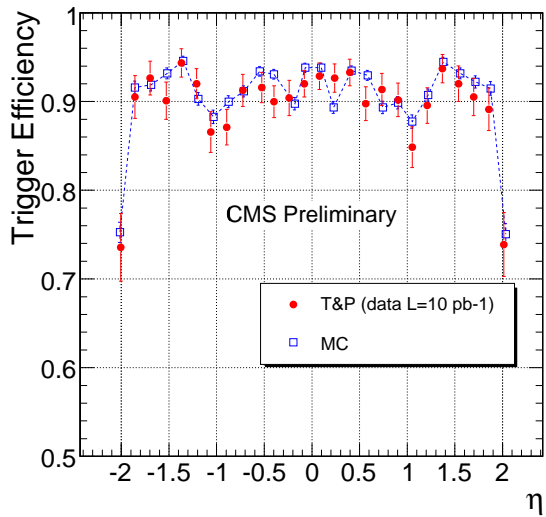}
\caption{The trigger efficiency for high-$p_T$ muons ($p_T >$ 20 GeV) as a function of $\eta$.} \label{tapmuon}
\end{figure*}

The efficiency of trigger, lepton reconstruction and selection can be measured directly from data using 
the Tag and Probe method~\cite{TAP}. 
The method relies upon $Z \rightarrow ll$ decays to provide an unbiased, high-purity, lepton sample with
which to measure the efficiency of a particular cut or trigger. One of the leptons, the ``tag'', is
required to pass stringent lepton identification criteria whilst the other lepton, the ``probe'',
is only required to satisfy a set of criteria depending on the efficiency under study.\\
Fig.~\ref{tapmuon} shows the single muon trigger efficiencies measured. 
It is computed on a $Z \rightarrow \mu^+\mu^-$ sample corresponding to
an integrated luminosity of 10 pb$^{-1}$. The results are shown as a function of the muon pseudorapidity.
The good agreement observed between the measured reconstruction efficiencies
using Tag and Probe method and the efficiencies evaluated using the Monte Carlo generator level
information is interpreted as a validation of the method.

\section{BACKGROUND STUDIES}
Electroweak background in $W \rightarrow l \nu$ is small and can be estimated with adequate precision from simulation. 
On the other hand the QCD backgroud is hard to estimate and control from simulation and therefore must be measured from the data. 
A missing transverse energy (transverse mass) template for the background is obtained requiring the full set 
of selection criteria but reverting the $\sigma_{\eta\eta}$ (muon isolation) one in the electron (muon) channel.
To obtain the signal missing transverse energy (transverse mass) template, we use $Z \rightarrow ll$ candidate removing one lepton 
to emulate the neutrino \cite{PASele} \cite{PASmuon}.

\end{document}